\documentclass[11pt]{article}  
\usepackage{amssymb}

\newcommand{\nc}{\newcommand}
\nc{\bb}{\begin{equation}} \nc{\ee}{\end{equation}}
\nc{\qq}{\qquad\qquad} \nc{\erm}{{\rm e}} \nc{\munu}{{\mu\nu}}
\nc{\dis}{\displaystyle} \nc{\um}{{1\over 2}}
\nc{\vecna}{\mbox{\boldmath $\nabla$}}
\nc{\ug}{\; = \;} \nc{\vs}{\vspace*}
\nc{\Hc}{{\cal H}}  \nc{\Lc}{{\cal L}}  \nc{\Lcn}{{\cal L}^{(n)}}
\nc{\Lcuno}{{\cal L}^{(1)}}  \nc{\Lcdue}{{\cal L}^{(2)}}
\nc{\Lctre}{{\cal L}^{(3)}}  \nc{\Lczero}{{\cal L}^{(0)}}
\nc{\Ebf}{\mbox{\boldmath $E$}} \nc{\Hbf}{\mbox{\boldmath $H$}}
\nc{\Vbf}{\mbox{\boldmath $V$}} \nc{\Fbf}{\mbox{\boldmath $F$}}
\nc{\Wbf}{\mbox{\boldmath $W$}} \nc{\lbf}{\mbox{\boldmath $l$}}
\nc{\xbf}{\mbox{\boldmath $x$}} \nc{\ubf}{\mbox{\boldmath $u$}}
\nc{\vbf}{\mbox{\boldmath $v$}} \nc{\wbf}{\mbox{\boldmath $w$}}
\nc{\jbf}{\mbox{\boldmath $j$}} \nc{\mubf}{\mbox{\boldmath $\mu$}}
\nc{\sigbf}{\mbox{\boldmath $\sigma$}}
\nc{\abf}{\mbox{\boldmath $a$}} \nc{\bbf}{\mbox{\boldmath $b$}}
\nc{\sbf}{\mbox{\boldmath $s$}} \nc{\dbf}{\mbox{\boldmath $d$}}
\nc{\rbf}{\mbox{\boldmath $r$}} \nc{\kbf}{\mbox{\boldmath $k$}}
\nc{\Lbf}{\mbox{\boldmath $L$}} \nc{\imp}{\mbox{\boldmath $p$}}
\nc{\albf}{\mbox{\boldmath $\alpha$}}
\nc{\dddov}{{\stackrel{\ldots}{v}}} \nc{\dox}{\dot{x}} \nc{\ddox}{\ddot{x}}
\nc{\dddox}{{\stackrel{\ldots}{x}}} \nc{\dopi}{\dot{\pi}} \nc{\dop}{\dot{p}}
\nc{\dov}{\dot{v}} \nc{\ddov}{\ddot{v}} \nc{\doa}{\dot{a}} \nc{\ddoa}{\ddot{a}}
\nc{\dddoa}{{\stackrel{\ldots}{a}}} \nc{\ddddoa}{{\stackrel{....}{a}}}
\nc{\ddddov}{{\stackrel{....}{v}}}
\nc{\Omn}{\Omega^\munu} \nc{\po}{\widehat{p}}
\nc{\rd}{{\rm d}} \nc{\dtau}{{\rd\tau}} \nc{\dt}{{\rd t}}
\nc{\pp}{{p_\mu v^\mu}} \nc{\ppo}{{\po_\mu \gamma^\mu}}
\nc{\ovsbf}{{\overline{\sbf}}}
\nc{\cmf}{_{\star}} \nc{\para}{^{\parallel}} \nc{\orto}{^{\perp}}
\nc{\vi}{{v^{(\rm i)}}} \nc{\aii}{{a^{(\rm 2i)}}}
\nc{\M}{{\rm I\!\!M}} \nc{\ain}{{a^{(\rm 2n)}}}
\nc{\impo}{\widehat{\imp}}  \nc{\Ho}{\widehat{H}}
\nc{\doS}{\dot{S}} \nc{\ddoS}{\ddot{S}} \nc{\doJ}{\dot{J}}
\nc{\doL}{\dot{L}} \nc{\ddoL}{\ddot{L}}
\nc{\doabf}{\mbox{\boldmath ${{\stackrel{.}{a}}}$}}
\nc{\ddoabf}{\mbox{\boldmath ${{\stackrel{..}{a}}}$}}
\nc{\dddoabf}{\mbox{\boldmath $\dddoa$}}
\nc{\ddddoabf}{\mbox{\boldmath $\ddddoa$}}
\nc{\CoMF}{_{\rm {\footnotesize CMF}}}
\nc{\drm}{{\rm d}}
\nc{\omz}{\omega_0}
\nc{\LcN}{{\cal L}^{(N)}}
\nc{\kb}{\overline{k}}
\nc{\alpt}{{\widetilde{\alpha}}}
\nc{\alm}{{\alpha_m^\mu}}
\nc{\alt}{{\alpt_m^\mu}}
\nc{\almd}{{\alpha^{\dagger\mu}_m}}
\nc{\altd}{{\alpt^{\dagger\mu}_m}}
\nc{\almbf}{{\albf_m}}
\nc{\altbf}{{{\widetilde{\albf}}_m}}
\nc{\almdbf}{{\albf^{\dagger}_m}}
\nc{\altdbf}{{\widetilde{\albf}^{\dagger}_m}}
\nc{\dosbf}{\mbox{\boldmath ${{\stackrel{.}{s}}}$}}
\nc{\lt}{\hat{l}}
\nc{\Ec}{E}  
\nc{\pa}{\partial}
\nc{\omf}{{\frac{\omega}{\omz}}}
\nc{\cost}{{\rm constant}}

\begin{document}

\title{DERIVING SPIN OF THE BOSONIC STRING$^{\star}$\\
\small{\centerline{[Published in {\em Found.Phys.Lett.}\,{\bf 19}, 367 (2006)]}}}

\footnotetext{$\!\!\!\!^{\star}\,$Work partially supported by I.N.F.N. and M.I.U.R.}

\author{GIOVANNI \ SALESI}
\date{}
\maketitle
\begin{center}
{{\em Universit\`a Statale di Bergamo, Facolt\`a di
Ingegneria,$^{\star\star}$\\
viale Marconi 5, 24044 Dalmine (BG), Italy}\\
and\\
{\em Istituto Nazionale di Fisica Nucleare--Sezione di Milano, Italy\\
via Celoria 16, 20133 Milan, Italy}}
\end{center}
\footnotetext{$\!\!\!\!^{\star\star}\,$e-mail: {\em salesi@unibg.it}}

\vs{0.5 cm}

\begin{abstract}
\noindent Exploiting the strict analogy between the motion of strings and
extended-like spinning particles, we propose an original kinematical formulation 
of the spin of bosonic strings and give, for the first time, an
analytical derivation of an explicit expression of the string spin vector.

\



\end{abstract}

\

\section{Proper-time equation of motion for a bosonic string}

By requiring the reparametrization and conformal invariances, the
Polyakov Lagrangian describing the free motion of a bosonic string can be written
in the conformal gauge\cite{Bailin, West, Polyakov, Green, Polchinski}
($\dis \dox\equiv\pa x/\pa\tau$, \ $\dis x^\prime\equiv\pa x/\pa\sigma$)
\bb
\Lc=\um M\left(\dox^2 - {x^{\prime}}^2\right)\,,
\ee
where the position $x(\tau, \sigma)$ at the proper time $\tau$ is
parameterized by $\sigma$. We have $M\dis\equiv L\cal T$, where $L\equiv
2\pi/\omz$ and $\cal T$ are the string ``length'' and the string tension, 
respectively\footnote{In the literature it is often assumed $\omz=2$,
$M=\pi\cal T$, $L=\pi$, with $\omz$, $\tau$ and $\sigma$ dimensionless.
In this paper, having to perform a harmonic analysis in the
$\tau$-sector, we prefer to assume time dimensions for $\sigma$ and $\tau$,
with $[\omz]=[t]^{-1}$.}.
The Euler-Lagrange equation is ($\mu=0,1,2,\dots,N$, where $N$ is the dimension of the
embedding space)
\bb
\ddox_\mu \ug x_\mu^{\prime\prime}\,.          \label{EL}
\ee
Considering a {\em closed} string\footnote{Equations and results quite 
analogous to the ones found in the present paper hold also for {\em open} strings.}
we have to require the following boundary condition 
\bb
x^\mu(\tau, \sigma) \ug x^\mu(\tau, \sigma+L)\,.   \label{boundary}
\ee
The general solution of Eq.\,(\ref{EL}) satisfying constraint
(\ref{boundary}) is ($c=1$)
$$
x^\mu(\tau,\sigma) = x_0^\mu + \frac{p^\mu}{M}\,\tau \ +
$$
\bb
+\,i\sqrt{\frac{\hbar}{2M\omz}}\sum_{m=1}^\infty
\frac{1}{m}\left[\alm\erm^{-im\omz(\tau-\sigma)}
+\alt\erm^{-im\omz(\tau+\sigma)}
+\almd\erm^{im\omz(\tau-\sigma)}
+\altd\erm^{im\omz(\tau+\sigma)}
\right]\,.                 \label{solgen}
\ee
Quantity $p^\mu$ indicates as usual the constant {\em mean} 
momentum (whose spatial part $\imp$, equal to $M$ times the 
center-of-mass velocity, vanishes in the center-of-mass frame) which is 
different from the (non-constant) total {\em canonical} momentum $P^\mu$ 
below defined. The dimensionless Fock operators $\alm$, $\alt$
obey the usual bosonic commutation rules.
Let us define the operators $a_m^\mu$, $\widetilde{a}_m^\mu$ according to
\bb
a_m^\mu \equiv \frac{1}{\sqrt{m}}\alm \qquad\qquad
\widetilde{a}_m^\mu \equiv \frac{1}{\sqrt{m}}\alt\,,  \label{amudef}
\ee
and impose the usual (equal-$\tau$) canonical commutation relations
$$
[x^\mu(\tau, \sigma), \ x^\nu(\tau, \sigma)] \ug
[v^\mu(\tau, \sigma), \ v^\nu(\tau, \sigma)] \ug 0\,,
$$
$$
[x^\mu(\tau, \sigma), \ P^\nu(\tau, \sigma)] \ug -i\hbar\delta(\sigma-
\sigma^\prime)g^\munu\,,
$$
where $\dis P^\mu=\frac{\pa\Lc}{\pa v_\mu}=Mv^\mu$ is the
canonical momentum conjugate to $x^\mu$.
As a consequence we obtain for $a^\mu$, $\widetilde{a}^\mu$ standard
harmonic oscillator commutators
\bb
[a_m^\mu, \ a_n^{\dagger\nu}] \ug [\widetilde{a}_m^\mu, \
\widetilde{a}_n^{\dagger\nu}] \ug -\delta_{m+n,0}\,g^\munu\,.
\ee
Differentiating Eq.\,(\ref{solgen}) with respect to $\tau$  
we get the expression of the velocity
\bb
v^\mu(\tau,\sigma) = \frac{p^\mu}{M} +
\sqrt{\frac{\hbar\omz}{2M}}\sum_{m=1}^\infty
\left[\alm\erm^{-im\omz(\tau-\sigma)}
+\alt\erm^{-im\omz(\tau+\sigma)}
-\almd\erm^{im\omz(\tau-\sigma)}
-\altd\erm^{im\omz(\tau+\sigma)}
\right]\,.                  \label{vmusol}
\ee
With respect to the variable $\tau$ the velocity is periodic with
period $\dis T=\frac{2\pi}{\omz}$: then we have \ $\forall\tau$,
$\forall\sigma$
$$
v_\mu\left(\tau+T,\sigma\right)-v_\mu\left(\tau-T,
\sigma\right)\ug 0\,.
$$
Expanding in Taylor series (the notation $^{(n)}$ indicates the
$n-$th $\tau-$derivative)
$$
\sum_{n=0}^\infty\frac{v_\mu^{(n)}(\tau,\sigma)}{n!}\,T^n \,-\,
\sum_{n=0}^\infty(-1)^n\frac{v_\mu^{(n)}(\tau,\sigma)}{n!}\,T^n \ug 0
$$
and then
\bb
\sum_{n=0}^\infty\frac{v_\mu^{(2n+1)}(\tau,\sigma)}{(2n+1)!}\,
T^{2n+1} \ug 0\,.  \label{veldiff}
\ee
By introducing, for convenience, coefficients $k_n$ defined as
\bb
k_n \equiv M\frac{(-1)^n}{(2n+1)!}T^{2n} =
M\frac{(-1)^n}{(2n+1)!}\left(\frac{2\pi}{\omz}\right)^{2n}\,,
\label{kndef}
\ee
we can re-write eq.\,(\ref{veldiff}) as follows
\bb
\sum_{n=0}^{\infty}(-1)^nk_nv_\mu^{(2n+1)} \ug 0\,. 
\label{proptimeq}
\ee
Time-integrating both sides of the above equation we get
\bb
\sum_{n=0}^{\infty}(-1)^nk_nv_\mu^{(2n)} \ug \cost\,,
\ee
which, being $k_0=M$ for (\ref{kndef}), can be re-written as
\bb
Mv = \cost + \sum_{n=1}^{\infty}(-1)^nk_nv_\mu^{(2n)}\,.
\label{cost}
\ee
By comparison between Eqs.\,(\ref{cost}) and (\ref{vmusol})
we infer that the constant is just the mean momentum
$p^\mu$. On the other hand inserting the expression of the velocity
given by (\ref{vmusol}) in the above equation and exploiting
definition (\ref{kndef}), we get, after some algebra, just constant=$p^\mu$:
So that we finally get a ``proper-time equation of motion'' for
a bosonic string in the form
\bb
\fbox{${\dis\sum_{n=0}^{\infty}(-1)^nk_nv_\mu^{(2n)} \ug p_\mu\,.}$} 
\label{harmonic}
\ee
For various applications in Section 3, let us introduce the dimensionless coefficients
$\kb_n$ defined as follows
\bb
\kb_n \equiv k_n\frac{\omz^{2n}}{M}\,.
\ee
Exploiting the explicit expression of $\kb_n$ we have for $x\neq 0$
$$
\sum_{n=0}^\infty \kb_n x^{2n} =
\frac{1}{2\pi x}\sum_{n=0}^\infty\frac{(-1)^n(2\pi)^{2n+1}}{(2n+1)!}x^{2n+1}
= \frac{\sin(2\pi x)}{2\pi x}\,.
$$
Differentiating side by side the previous equation
we get
$$
\sum_{n=0}^\infty n\kb_n x^{2n} =
\frac{\cos(2\pi x)}{2}-\frac{\sin(2\pi x)}{4\pi x}\,,
$$
which for $x=m\in\mathbb{N}^+$ yields
\bb
\sum_{n=0}^\infty n\kb_n m^{2n} \ug 
\frac{1}{2}\,.                  \label{sumpesata}
\ee
We have also    
\bb
\sum_{n=0}^\infty\kb_n m^{2n} \ug 0     \label{sumkbm}
\ee
holding for any positive integer $m$: this relation can be obtained 
taking $x$ equal to $m\in\mathbb{N}^+$ in
$$
\sum_{n=0}^\infty\kb_n x^{2n} = \frac{\sin(2\pi x)}{2\pi x}\,.
$$

\

\section{Non-Newtonian mechanics: a short review}

In recent papers \cite{NNM} we proposed and developed a classical
symplectic theory for extended-like\footnote{The term {\em extended-like}
refers to spinning systems which, even if not ``materially'' extended as
strings, nevertheless are something halfway between a point and a
rotating rigid body (as, e.g., a top). In fact the center of mass and the
center of charge are distinct points, velocity and momentum are not
parallel, and we observe an internal microscopic motion besides the
macroscopic external one (the so-called ``Zitterbewegung''\cite{ZBW,Salesi,Schroedinger}).}
microsystems accounting for spin and Zitterbewegung.
The classical motion of spinning particles was therein described without
recourse to particular models or special formalisms, and without employing
Grassmann variables, Clifford algebras, or classical spinors, but simply by
generalizing the standard spinless theory.
It was only assumed the invariance with respect to the Poincar\'e group;
and only requiring the conservation of the linear and angular momenta we
derived the Zitterbewegung and the other kinematical properties and motion constraints.
Newtonian Mechanics is re-obtained as a particular case of that theory:
namely for spinless systems with no Zitterbewegung.

We started with a Poincar\'e-invariant Lagrangian which generalizes the
Newtonian Lagrangian $\Lczero=\um mv^2$ (as usual $v^2\equiv v_\mu
v^ \mu$) by means of proper-time derivatives of the velocity up to the $N$-th
order (we take the scalar potential $U=0$ considering only free particles)
\bb
\LcN \equiv \um Mv^2 + \um k_1\dov^2 + \um k_2\ddov^2 + \cdots -
\equiv \sum_{n=0}^N\,\um k_n^N{v^{(\rm n)}}^2\,,  \label{LcN}
\ee
where the $k_n^N$ are scalar coefficients analogous to the $k_n^N$ introduced in Section 1.

The coefficients $k_i$ ---which may be chosen equal to zero for $i$ larger than
a given integer--- might be fixed by the self-interaction of the particle, and are often
functions of mass and charge. Let us recall, for comparison, the infinite-terms equation 
of the self-radiating classical electron, Caldirola's ``chronon'' theory of the 
electron\cite{Caldirola}, the so-called ``dressed particles'' described in \cite{Korsyakov}. 
In other theoretical frameworks, the coefficients $k_i$ can be related to the underlying geometrical
structure of spinning-particles or strings (or $D$-branes). Some authors 
\cite{Pavsic,Plyushchay,Polyakov2} have proposed classical actions, describing ``rigid particles'' 
and ``rigid strings'', in which appear additional terms dependent on the so-called ``extrinsic curvature'', 
that is, on the 4-acceleration squared. 
Classical equations of the motion for a rigid particle or for a rigid $n-$dimensional worldsheet, 
either in flat or curved background spacetimes, have been derived from Lagrangians containing
also terms dependent on higher derivatives of the 4-velocity (``torsion''-terms,
etc.). In \cite{Nesterenko} the equations of motion are reformulated in terms of the 
principal wordline curvatures which turn out to be motion integrals, namely mass and
spin. Another interesting result in\cite{Nesterenko} is the emergence of a particle maximal
proper acceleration: which constitutes an example of a consistent relativistic 
dynamics obeying the principle of a superiorly limited value of the acceleration, recently
advanced. 
In a sense most of the above-mentioned models ---we refer to those approaches
where the time parameter, i.e., the proper time, is just the time in the center-of-mass frame--- 
are {\em particular cases} of the present theory: As a matter of fact we do not choose
a particular (internal) geometry or kinematics, but include a priori time derivatives of any order, 
leaving the structure-coefficients $k_i$ not fixed. 

Coming back to (\ref{LcN}), the Euler-Lagrange equation of motion
\bb
\frac{\pa\Lc}{\pa x} = \frac{\rd}{\dtau}\left({\frac{\pa\Lc}{\pa\dox}}\right) -
\frac{\rd^2}{\dtau^2}\left({\frac{\pa\Lc}{\pa\ddox}}\right) +
\frac{\rd^3}{\dtau^3}\left({\frac{\pa\Lc}{\pa\dddox}}\right) -
\cdots
\ee
gives a constant-coefficients $N$-th order differential equation, which
appears as a generalization of Newton's Law $F=Ma$, in which non-Newtonian
Zitterbewegung terms appear (having assumed $U=0$, here obviously $F=0$):
\bb
0 \ug Ma^\mu + \sum_{n=1}^N\,(-1)^{{\rm n}}k_n^N\,\aii^\mu\,. \label{GNEq}
\ee
The zero-th order canonical momentum
$\dis\frac{\pa\Lc}{\pa\dox_\mu} -
\frac{\rd}{\dtau}\left({\frac{\pa\Lc}{\pa\ddox_\mu}}\right) +
\frac{\rd^2}{\dtau^2}\left({\frac{\pa\Lc}{\pa\dddox_\mu}}\right) -
\cdots$ \ conjugate to $x^\mu$
writes
\bb
p_{[0]}^\mu =  Mv^\mu + \sum_{n=1}^N\,(-1)^{{\rm n}}\,k_n^N\,{v^{(\rm 2n)}}^\mu =
\sum_{n=0}^N\,(-1)^{{\rm n}}\,k_n^N\,{v^{(\rm 2n)}}^\mu\,.
\label{ZeroMom}
\ee
{\em This equation coincides with the proper time equation of motion for
a bosonic string Eq.\,}(\ref{harmonic}).
By satisfying the symmetry of the Lagrangian under 4-rotations, which
implies the conservation of the total angular momentum, the classical
spin can be unequivocally defined employing only classical kinematical
quantities:
Through the N\"other Theorem we derived the spin tensor and the spin vector, 
respectively
\bb
S_\munu \ug \sum_{n=1}^Nk_n^N\,\sum_{l=0}^{n-1}(-1)^{n-l-1}
\left(v_\mu^{(l)}v_\nu^{(2n-l-1)}-v_\nu^{(l)}v_\mu^{(2n-l-1)}\right)\,;
\ee
\bb
\sbf \ug \sum_{n=1}^Nk_n^N\,\sum_{l=0}^{n-1}(-1)^{n-l-1}
\vbf^{(l)}\times\vbf^{(2n-l-1)}\,.           \label{NNMspin}
\ee
As an example let us take $N=4$. We have
\bb
\imp \ug M\vbf-k_1\doabf+k_2\dddoabf-k_3\abf^{({\rm V})}+
k_4\abf^{({\rm VII})}\,.
\ee
The spin is
$$
\sbf \ug k_1(\vbf\times\abf) + k_2(\abf\times\doabf-\vbf\times\ddoabf)
+k_3\left(\doabf\times\ddoabf-\abf\times\dddoabf+
\vbf\times\abf^{({\rm IV})}\right)+
$$
\bb
+k_4\left(\ddoabf\times\dddoabf-\doabf\times\abf^{({\rm IV})}+
\abf\times\abf^{({\rm V})}- \vbf\times\abf^{({\rm VI})}\right)\,;
\ee
thus, after differentiating and simplifying,
$$
\dosbf \ug \vbf\times\left(k_1\doabf-k_2\dddoabf+k_3\abf^{({\rm V})}-
k_4\abf^{({\rm VII})}\right) = \vbf\times(M\vbf-\imp)=\imp\times\vbf\,,
$$
as expected [see Eq.\,(\ref{dosbf}) in the next section].

The Hamiltonian representation of the theory is performed by
introducing, besides the (constant) zero-th order momentum $p_{[0]}^\mu$ given in
(\ref{ZeroMom}), the other (non-constant) $l$-th order momenta $p_{[l]}^\mu$
canonically conjugate to $x_{[l]}\equiv x^{(l)}$:
\bb
p_{[l]}^\mu \equiv \sum_{n=l}^N(-1)^{n-l}\frac{\rd^{n-l}}{\dtau^{n-l}}
\left(\frac{\pa\Lc}{\pa x^{(n+1)}}\right)
= \sum_{n=l}^N(-1)^{n-l}k_n^Nv^{(2n-l)}
\ee
[this definition contains also the $l=0$ case, Eq.\,(\ref{ZeroMom})].
Employing the high order momenta the spin vector (\ref{NNMspin})
can be expressed also in a canonical way:
\bb
\sbf \ug \sum_{l=1}^N\xbf_{[l]}\times\imp_{[l]}\,.
\ee
analogous to the orbital angular momentum $\lbf=\xbf\times\imp_{[0]}$.
The {\em conserved} scalar Hamiltonian, obtained imposing the
$\tau$-reparametrization invariance of the Lagrangian, is
\bb
\Hc = \sum_{l=0}^Np_{[l]}^\mu\dox^{(l)}_\mu - \Lc =
\um Mv^2 + \sum_{n=1}^Nk_n^N\,\left[\um{v^{(n)}}^2 +
\sum_{l=0}^{n-1}(-1)^{n-l}v^{(l)\mu}v^{(2n-l)}_\mu\right]\,.
\label{NNMscalar}
\ee
It can be also shown that a couple of Hamilton equations
\bb
\dox_{[l]}^\mu \ug \frac{\pa\Hc}{\pa p_{[l]\mu}}  \qquad\qquad\quad
\dop_{[l]}^\mu \ug -\frac{\pa\Hc}{\pa x_{[l]\mu}}
\ee
holds for any couple of canonical variables $\dis\left(x_{[l]}^\mu;
\ p_{[l]}^\mu\right)$, and that the set of the Hamilton equations is
globally equivalent to the Euler-Lagrange equation (\ref{GNEq}).

\

\section{String motion constants}

Let us turn back to bosonic strings.
For a free spinning microsystem the angular momentum, defined as the sum
of the orbital angular momentum $x^\mu p^\nu - x^\nu p^\mu$ and of
the intrinsic angular momentum, conserves
\bb
\doJ^\munu = \doL^\munu + \doS^\munu = 0\,.
\ee
Since the momentum is constant, $\dop^\mu=0$, we get $\doL^\munu = v^\mu
p^\nu - v^\nu p^\mu$: therefore we obtain
\bb
\doS^\munu \ug p^\mu v^\nu - p^\nu v^\mu\,.
\ee
As a consequence, the equation of the motion for the spin 3-vector writes
\bb
\dosbf \ug \imp\times\vbf\,;           \label{dosbf}
\ee
and in the center-of-mass frame where $\imp=0$ we have, as expected, a constant
spin, \ $\dosbf = 0.$ \
Besides the trivial choice $\sbf=-\Lbf=\imp\times\xbf$,
the only vector $\sbf(\tau; \sigma)$ which can be builted up with the quantities at
our disposal (that is $\xbf$ and its derivatives) and satisfies (\ref{dosbf}) 
has the same formal expression (\ref{NNMspin}) for the spin of an extended-like 
particle
\bb
\sbf \ug \sum_{n=1}^\infty k_n\,\sum_{l=0}^{n-1}(-1)^{n-l-1}
\vbf^{(l)}\times\vbf^{(2n-l-1)}\,.           \label{spin}
\ee
This statement can be immediately realized taking into account
that {\em both the bosonic string and the extended-like particle obey
the same proper time equation of motion}, Eq.\,(\ref{harmonic}),
provided that coefficients $k_n$ be defined according to
Eq.\,(\ref{kndef}).
Alternatively we can start time-differentiating side by side the previous
equation
$$
\dosbf = \sum_{n=1}^\infty k_n\,\sum_{l=0}^{n-1}(-1)^{n-l-1}
\left[\vbf^{(l+1)}\times\vbf^{(2n-l-1)}+\vbf^{(l)}\times\vbf^{(2n-l)}\right]\,.
$$
Putting $\lt\equiv l+1$ we have
$$
\dosbf = \sum_{n=1}^\infty k_n\,
\left[\sum_{\lt=1}^n(-1)^{n-\lt}\vbf^{(\lt)}\times\vbf^{(2n-\lt)} +
\sum_{l=0}^{n-1}(-1)^{n-l-1}\vbf^{(l)}\times\vbf^{(2n-l)}\right]=
$$
$$
= \sum_{n=1}^\infty k_n\,
\left[\sum_{\lt=1}^{n-1}(-1)^{n-\lt}\vbf^{(\lt)}\times\vbf^{(2n-\lt)} +
\sum_{l=1}^{n-1}(-1)^{n-l-1}\vbf^{(l)}\times\vbf^{(2n-l)}\right. +
$$
$$
\left. + \vbf^{(n)}\times\vbf^{(n)}  + (-1)^{n-1}\vbf\times\vbf^{(2n)}
\right]\,.
$$
In the above sum the first two terms mutually cancel,
the third one is identically zero; the only contribute comes from the last
term:
$$
\dosbf = - \vbf\times\sum_{n=1}^\infty(-1)^nk_n\,\vbf^{(2n)}\,.
$$
Then, because of (\ref{harmonic}), we have
$$
\dosbf = \imp\times\vbf\,.
$$

\

\noindent By quite analogous arguments it can be proved that, not
only for extended-like particles, but also for bosonic strings,
the quantity appearing in Eq.\,(\ref{NNMscalar})
\bb
\Ec \ug \um Mv^2 + \sum_{n=1}^\infty k_n\,\left[\um{v^{(n)}}^2 +
\sum_{l=0}^{n-1}(-1)^{n-l}v^{(l)\mu}v^{(2n-l)}_\mu\right]  \label{scalar}
\ee
is the only 4-scalar integral of motion  
which can be builted trough the quantities at our disposal ($\imp$,
$\xbf$ and its derivatives).    
Therefore, for any solution (\ref{vmusol}) it is
\bb
\dot{\Ec} \ug 0\,.
\ee
For a bosonic string, because of Eqs.\,(\ref{kndef}),
(\ref{spin}), (\ref{scalar}), we have the following expressions for the spin
and $\Ec$, respectively:
\bb
\sbf \ug M\sum_{n=1}^\infty\frac{1}{(2n+1)!}
\left(\frac{2\pi}{\omz}\right)^{\!\!2n}
\,\sum_{l=0}^{n-1}(-1)^{l+1} \vbf^{(l)}\times\vbf^{(2n-l-1)}
\label{spininf}
\ee
\bb
\Ec \ug \um Mv^2 +
\sum_{n=1}^\infty\frac{(-1)^n}{(2n+1)!}\left(\frac{2\pi}{\omz}\right)^{\!\!2n}
\,\left[\um{v^{(n)}}^2 +
\sum_{l=0}^{n-1}(-1)^{n-l}v^{(l)\mu}v^{(2n-l)}_\mu\right] \ .
\label{scalarinf}
\ee

\

\

\noindent Let us choose the center-of-mass frame ($\imp=0$) as reference frame, and
insert the solution (\ref{vmusol}) with $\imp=0$ in (\ref{spin}).
After some algebra we are able to represent the ($\tau-$constant in the 
center-of-mass frame) spin vector in terms of operators (\ref{amudef})
$$
\sbf(\sigma) \ug \hbar\sum_{m=1}^\infty\left(\sum_{n=1}^\infty
n\kb_n m^{2n}\right)i\left[\left(\abf_m^\dagger\times\abf_m\right) +
\left(\widetilde{\abf}_m^\dagger\times\widetilde{\abf}_m\right) +
\right.
$$
$$
\left. + \; \left(\abf^\dagger_m\times\widetilde{\abf}_m\right)\,\erm^{-2im\omz\sigma} +
\left(\widetilde{\abf}_m^\dagger\times\abf_m\right)
\,\erm^{2im\omz\sigma}\right]\,,
$$
which, accounting for (\ref{sumpesata}), reduces to
\bb
\sbf(\sigma) = \hbar\sum_{m=1}^\infty\frac{i}{2}
\left[\left(\abf_m^\dagger\times\abf_m\right) +
\left(\widetilde{\abf}_m^\dagger\times\widetilde{\abf}_m\right) +
\left(\abf^\dagger_m\times\widetilde{\abf}_m\right)\,\erm^{-2im\omz\sigma} +
\left(\widetilde{\abf}_m^\dagger\times\abf_m\right)
\,\erm^{2im\omz\sigma}\right]\,.
\ee
Thus the spin averaged over a closed string results to be
\bb
\fbox{${\dis\overline{\sbf} = \frac{1}{L}\int_0^L\sbf(\sigma)\rd\sigma
= \hbar\sum_{m=1}^\infty\frac{i}{2}
\left[\left(\abf^\dagger_m\times\abf_m\right) +
\left(\widetilde{\abf}^\dagger_m\times\widetilde{\abf}_m\right)\right]}$} \ ,
\label{Fourier}
\ee
which agrees with the usual expression (not explicitly derived but only 
postulated) in standard string theory.

\

\noindent Finally, inserting (\ref{vmusol}) in (\ref{scalar}), 
after $\sigma-$integration 
we get
$$
\overline{\Ec} = \frac{1}{L}\int_0^L\Ec(\sigma)\rd\sigma =
$$
$$
= \frac{p^2}{2M} - \frac{\hbar\omz}{4}\sum_{m=1}^\infty
\frac{1}{m}\left(1+4\sum_{n=1}^\infty n\kb_n m^{2n}\right)
\left[\left(a_{m\mu}^\dagger a_m^\mu+a_{m\mu}a_m^{\dagger\mu}\right)+
\left(\widetilde{a}_{m\mu}^\dagger\widetilde{a}_m^\mu+
\widetilde{a}_{m\mu}\widetilde{a}_m^{\dagger\mu}\right)\right]\,;
$$
or, exploiting (\ref{sumkbm}) and (\ref{sumpesata}),
\bb
\overline{\Ec} \ug \frac{p^2}{2M} -
\frac{\hbar\omz}{2}\sum_{m=1}^\infty\frac{1}{m}
\left[\left(a_{m\mu}^\dagger a_m^\mu+a_{m\mu}a_m^{\dagger\mu}\right) +
\left(\widetilde{a}_{m\mu}^\dagger\widetilde{a}_m^\mu+
\widetilde{a}_{m\mu}\widetilde{a}_m^{\dagger\mu}\right)\right]\,.
\ee
Taking into account that $M=2\pi{\cal T}/\omz$ we can put the
above equation in the form
\bb
\fbox{${\dis\overline{\Ec} \ug \frac{p^2}{2M} -
\frac{h\cal T}{2M}\sum_{m=1}^\infty\frac{1}{m}
\left[\left(a_{m\mu}^\dagger a_m^\mu+a_{m\mu}a_m^{\dagger\mu}\right) +
\left(\widetilde{a}_{m\mu}^\dagger\widetilde{a}_m^\mu+
\widetilde{a}_{m\mu}\widetilde{a}_m^{\dagger\mu}\right)\right]}$} \ ,
\label{E-integrated}
\ee
where we see that the quantum (due to the ``internal'' string motion)
contribution to $\overline{\Ec}$ grows in modulus with the tension for mass unity
${\cal T}/M$.

\noindent The conserved quantity $\Ec(\sigma)$ given by (\ref{scalarinf}) can be 
put in comparison with the {\em non-conserved}\footnote{Actually if we insert the general 
solution of the string motion equation given by Eq.(\ref{solgen}) in the standard 
expression Eq.(\ref{Hstandard}), we do {\em not} get a time-constant expression; as 
instead it is obtained if we insert the solution in Eq.(\ref{scalarinf}).} Hamiltonian 
density found in the literature on bosonic string\cite{Bailin, West, Polyakov, Green, Polchinski}
\bb
\Hc(\tau;\sigma) = P_\mu\dox^\mu - \Lc =
\um M\left(\dox^2 + {x^{\prime}}^2\right)\,, \label{Hstandard}
\ee
which anyhow, taking the string-average, gives back just Eq.\,(\ref{E-integrated})
(which is conserved):
\bb
\overline{\Hc}=\frac{1}{L}\int_0^L\Hc\,\rd\sigma \ug \overline{\Ec}\,.
\ee

\

\section{Conclusions}
In this paper we have analytically derived the spin vector of bosonic strings 
obtaining an explicit kinematical formulation of the intrinsic angular momentum 
through a sum over all string modes. 
We have exploited the common {\em non-local} character of strings and spinning 
extended-like particles. Actually one of the novelties of the present approach
is describing a bosonic string trough infinite time derivatives: 
the spin of the string arises as angular momentum of such a mechanical system 
in the center-of-mass frame.

The present derivation of the bosonic string spin through a Fourier expansion as in
({\ref{Fourier}), or in a pure kinematical form as in (\ref{spin}), 
is, as far as we know, original and unpublished in the literature on strings. 
 
\

\

{\bf Acknowledgements}

\noindent The author is grateful to L.\,Brandolini, G.\,Cavagna, E.\,Di Grezia,
S.\,Esposito for mathematical suggestions. The scientific collaboration of 
F.\,Bottacin, G.\,Gigante, E.\,Recami, M.\,Villa and D.\,Zappal\`a is also acknowledged.

\

\

\

\end{document}